\documentclass[useAMS,usenatbib]{mn2e}

\usepackage{graphicx}
\usepackage{amsmath}
\usepackage{xspace}
\usepackage{dsfont}

\title[]
{Fast Bayesian Inference for Exoplanet Discovery in Radial Velocity Data}
    
\author[Brewer and Donovan]{%
  Brendon~J.~Brewer$^{1}$\thanks{bj.brewer@auckland.ac.nz},
  Courtney P. Donovan$^{1}$
  \medskip\\
  $^1$Department of Statistics, The University of Auckland, Private Bag 92019, Auckland 1142, New Zealand}

\begin{document}
             
\date{To be submitted to MNRAS}
             
\maketitle

\label{firstpage}


\begin{abstract}
Inferring the number of planets $N$ in an exoplanetary system from radial velocity
(RV) data is a challenging task. Recently, it has become clear that RV data
can contain periodic signals due to stellar activity, which can be difficult
to distinguish from planetary signals. However, even doing the inference
under a given set of simplifying assumptions (e.g. no stellar activity) can
be difficult. It is common for the posterior distribution for the
planet parameters, such as orbital periods,
to be multimodal and to have other awkward features. In
addition, when $N$ is unknown, the marginal likelihood (or evidence) as a
function of $N$ is required. Rather than doing separate runs with different
trial values of $N$, we propose an alternative
approach using a trans-dimensional Markov Chain Monte Carlo method within
Nested Sampling. The posterior distribution for $N$ can be obtained with a
single run. We apply the method to $\nu$ Oph and Gliese 581, finding moderate evidence
for additional signals in
$\nu$ Oph with periods of 36.11 $\pm$ 0.034 days,
75.58 $\pm$ 0.80 days, and 1709 $\pm$ 183 days;
the posterior probability that at least one of these
exists is 85\%. The results also suggest Gliese 581 hosts many (7-15) ``planets''
(or other causes of other periodic
signals), but only 4-6 have well determined periods. The analysis of both
of these datasets shows phase transitions exist which are difficult to
negotiate without Nested Sampling.
\end{abstract}

\begin{keywords}
stars: planetary systems --- techniques: radial velocities ---
methods: data analysis --- methods: statistical
\end{keywords}


\section{Introduction}
The number of known extrasolar planets has exploded in the last two
decades. This has been driven by improvements in
all of the different techniques used to detect and characterise exoplanets,
including the radial velocity (RV) method \citep[e.g.][]{2012PASJ...64..135S},
the transit method \citep[e.g.][]{2014PNAS..11112647B},
and gravitational microlensing
\citep[e.g.][]{2014ApJ...785..155B, 2014ApJ...790...14Y}.

The problem of inferring the properties of an
exoplanetary system from observational data can be challenging.
In the case of radial velocity data,
the expected signal due to an exoplanet is periodic, and the goal is to
infer the number of planets in the system, as well as their properties such
as orbital periods and eccentricities. Many
different techniques have been proposed for doing this.
These techniques fall into two main
classes: i) those based on periodograms \citep[e.g.][]{2009A&A...496..577Z},
and ii) those based on model fitting in the Bayesian inference framework,
to describe the uncertainties probabilistically
\citep[e.g.][]{2011MNRAS.410...94G, 2014MNRAS.437.3540F,
2011A&A...528L...5T, fengji}. Bayesian model fitting via Markov Chain Monte
Carlo (MCMC) tends to be computationally intensive, especially if we
want to calculate the posterior distribution for $N$, the number of planets.

It is well known that RV datasets can contain periodic signals
resulting from stellar activity rather than planets,
which can affect the conclusions we draw about
exoplanet systems \citep[e.g.][]{2001A&A...379..279Q, 2007A&A...474..293B, 2014Sci...345..440R}.
Therefore, it is important to develop models which attempt to distinguish
stellar activity signals from Keplerian planet signals based on the shape of
the oscillations and/or additional data constraining the periods of any
stellar activity signals. We do not address this important challenge in the
present paper. Rather, we consider the problem of inferring the number $N$ of
Keplerian signals in an RV dataset in a computationally efficient way, under
the simplifying assumption that only Keplerian signals are present in the data.

We introduce a trans-dimensional birth-death MCMC approach
\citep{birthdeath} to inferring $N$.
When $N$ is treated as just another model parameter, we can obtain its
posterior distribution in a single run.
In addition, rather than trying to sample the posterior distribution, 
we use Diffusive Nested Sampling
\citep[DNS][]{dnest} which replaces the posterior distribution with an alternative
{\it mixture of constrained priors}, allowing mixing between separated modes.
As a result, we are able to sample the posterior distribution for $N$, and
evaluate the marginal likelihood (including the sum over $N$) in
a single run which takes about 10 minutes on a 2-3 planet system.
On the other hand the approach of
\citet{2011MNRAS.410...94G} takes approximately 30 minutes per planet
(Gregory, priv. comm.). A C++ implementation of our method is available online
at {\tt https://github.com/eggplantbren/Exoplanet} under the terms of the
GNU General Public Licence.

\section{Inference}
Bayesian inference is the use of probability theory to describe uncertainty
\citep{sivia, ohagan}. In this framework, we approach data analysis problems by first
constructing a {\it hypothesis space}, which is the set of possible answers
to the problem we are considering. Normally, this is the set of possible
values of a vector of parameters $\theta$ whose values we want
to know. We then assign probability distributions
called the {\it prior} and the {\it sampling distribution}. The prior
distribution $p(\theta)$ describes
our initial uncertainty about which values of the parameters
$\theta$ are plausible, and the
sampling distribution $p(D | \theta)$ describes
our initial uncertainty about the data set we're going to observe, as a
function of the unknown parameters $\theta$.
When the data is known, our state of knowledge about the parameter is
updated from the prior $p(\theta)$ to the posterior distribution given by
Bayes' rule:
\begin{eqnarray}
p(\theta | D) &=&
\frac{p(\theta)p(D | \theta)}
{p(D)}
\end{eqnarray}
where $p(D | \theta)$ as a function of $\theta$ is called the likelihood,
once the actual dataset has been substituted in. Note that some authors do not
distinguish between a
sampling distribution and a likelihood. Throughout this paper we use the term
sampling distribution for $p(D|\theta)$ if we are discussing a probability
distribution (actually a family of them, indexed by $\theta$)
over the set of possible datasets. We use the term {\it likelihood}
when the actual dataset has been plugged in, when $p(D|\theta)$
becomes a scalar function (not a probability distribution)
over the parameter space.

The denominator, often called the {\it evidence} or
{\it marginal likelihood}, is given by the expected value of the likelihood
with respect to the prior:
\begin{eqnarray}
\mathcal{Z} = p(D) = \int p(\theta) p(D | \theta) d^n \theta
\end{eqnarray}
where the integral is over the entire $n$-dimensional parameter space.
In the context of Bayesian computation, the prior is often denoted $\pi(\theta)$,
the likelihood $L(\theta)$, and the marginal likelihood $\mathcal{Z}$.

\subsection{Inferring the number of planets}
The number of orbiting planets, $N$, is an important parameter.
To calculate the posterior distribution for $N$, most authors
consider various trial values of $N$, and calculate the marginal likelihood
\begin{eqnarray}
p(D | N) &=& \int p(\theta | N) p(D | \theta, N) \, d^n \theta
\end{eqnarray}
for each possible value of $N$
\citep[e.g.][]{2011MNRAS.415.2523G, 2011MNRAS.415.3462F, 2014MNRAS.437.3540F, fengji}, marginalising
over all other model parameters.
The posterior distribution for $N$ can then be found straightforwardly by
using Bayes' rule with $N$ as the only unknown parameter:
\begin{eqnarray}
p(N | D) &=& \frac{p(N)p(D | N)}{\sum_N p(N)p(D | N)}.
\end{eqnarray}
Popular
methods for calculating the marginal likelihood are Nested Sampling
\citep{skilling} and ideas related to thermodynamic integration
\citep[e.g.][]{neal}. Relationships between these methods are discussed by
\citet{cameron} and~\citet{scott}.

This traditional approach can be very time consuming.
Methods for calculating the marginal likelihood are
already more intensive than standard MCMC methods for sampling the posterior,
because they usually involve a sequence of probability distributions
(e.g. the constrained priors in Nested Sampling, or the annealed distributions
in thermodynamic integration) rather than a single distribution (the posterior).
This intensive process needs to be run many times, for $N=0$, $N=1$, $N=2$, and
so on.

The traditional approach to inferring $N$ also contradicts
fundamental ideas in Bayesian
computation. Imagine we are trying to compute the posterior distribution for
a parameter $a$ in the presence of a nuisance parameter $b$. This is usually solved
by exploring the joint posterior for $a$ and $b$, and then only looking at the
generated values of $a$. Nobody would suggest the wasteful alternative
of using a discrete grid of possible $a$ values and doing an entire Nested
Sampling run for each, to get the marginal likelihood as a function of $a$.
When the hypothesis space for $a$ is discrete, MCMC is still possible and there
is no reason to switch to the wasteful alternative.

\subsection{Trans-dimensional MCMC}
Trans-dimensional MCMC methods such as birth-death MCMC \citep{birthdeath} or the
more general reversible jump MCMC \citep{green} treat the model dimension
$N$ as just another model parameter. At fixed $N$, standard techniques such
as the Metropolis algorithm can be used to explore the posterior distribution.
Additional moves that propose to change the value of $N$ are also defined. The
simplest of these are birth-death moves. More complicated moves, such as
split-and-merge, are possible but not always necessary.
Trans-dimensional MCMC is a natural tool for a wide range of astronomical
data analysis problems \citep[e.g.][]{umstatter, walmswell, starfield, 2014arXiv1411.7447J}.

In the exoplanet context, a birth move
proposes to add one more planet to the model. The new planet's properties
(period, amplitude, eccentricity, etc) are drawn from their prior distribution
which may depend on other other model parameters or hyperparameters.
The corresponding death move simply chooses a planet currently in the model,
and removes it. The acceptance probability for these moves is 1 if we want
to explore the prior. To implement these moves in Nested Sampling
\citep{skilling}, where the target distribution is proportional to the prior but with a hard likelihood
constraint, then the acceptance probability is 1 if the proposed move
satisfies the likelihood constraint, and 0 if it does not.

A recent paper \citep{rjobject} introduced a general approach to implementing
trans-dimensional models within Diffusive Nested Sampling \citep{dnest}, a
general MCMC algorithm. The \citet{rjobject} software predefines the
Metropolis proposals for exploring trans-dimensional target distributions,
including when the prior for the properties of each model component (i.e. each
planet) is defined hierarchically.

\subsection{Phase transitions}
It is well known that Bayesian computation (using MCMC for example) can be
difficult when the posterior distribution is multimodal or has strong
dependencies between parameters. An uncommon but less well-known difficulty
is the existence of {\it phase transitions} \citep{skilling}.

Imagine a high-dimensional unimodal posterior distribution that is composed of a
broad, high volume but low density ``slab'' with a narrow, low volume but high
density ``spike'' on top of it.
An example is a mixture of two concentric high-dimensional gaussians with
different widths. If you ran MCMC on such a posterior, it would be difficult
to jump between the slab and the spike components. If the MCMC is currently
in the spike region (or phase) it will be unable to escape: a proposed move
into the slab will be rejected because of the ratio of densities. Conversely,
if the MCMC was in the slab region, it would be unlikely to go into the spike
region, because its volume is so small: it would be very unlikely to
{\it propose} to move into the spike. Thus, the situation behaves much like
a multimodal posterior, despite only being unimodal.

If the slab contains a very small amount of posterior probability, it is not
a problem if an MCMC algorithm spends all its time in the spike. However, this
situation could still cause problems with the calculation of the marginal likelihood
if annealing methods are used. The thermodynamic integral formula gives
the log of the marginal likelihood $\mathcal{Z}$ as an average of log likelihoods:
\begin{eqnarray}
\log(\mathcal{Z}) &=& \int_0^1 \left<\log\left[L(\theta)\right]\right>_\beta \, d\beta
\end{eqnarray}
where the expectation is taken with respect to the
distribution with ``inverse temperature'' $\beta$, proportional to
$\pi(\theta)L(\theta)^\beta$. Even if the slab contains virtually zero probability
when $\beta=1$ (i.e. the posterior), for some values of the inverse temperature
$\beta$ the slab and the spike will both be important. At
these temperatures the MCMC will fail to mix (it will incorrectly spend all its
time in either the slab or the spike, rather than mixing between the two)
and will give a misleading estimate of the average log likelihood at that temperature
and therefore an incorrect marginal likelihood estimate.

Phase transitions are well known in statistical mechanics, but can also occur
in Bayesian data analysis. Typically this occors when the data contains
a ``big'' effect which provides a lot of information about some parameters, and
a ``small'' subtle effect as well.
Nested Sampling, and variants such as DNS, are not affected
by phase transitions because the exploration only makes use of likelihood
{\it rankings}, rather than likelihood values themselves, and are therefore
invariant under monotonic transformations of the likelihood function.
Part of their output, the relationship between the
likelihood $L$ and the enclosed prior mass
$X(L) = \int \pi(\theta) \mathds{1}\left[L(\theta) > L\right]\, d^n\theta$
can be used to diagnose whether the
problem contains a phase transition. In particular, if the graph of $\log(L)$
vs. $\log(X)$ is convex at some point, then a phase transition exists
\citep{skilling}.

\subsection{Parameters and Priors}
To fit a planet model to RV data, we need parameters to describe
the properties of each planet. For simplicity, we describe each planet by
five parameters: the orbital period $P_i$, the semi-amplitude (in metres
per second) of the RV signal $A_i$, the phase of the signal
$\phi_i$ (defined such that $\phi=0$ gives an RV signal whose maximum is at
$t=0$), the eccentricity $e_i$, and the ``viewing angle'' $\omega_i$
(also known as the longitude of the line of sight). We defined our parameters
such that in the limit of zero eccentricity, the RV signal of a planet
reduces to $A_i\cos\left(2\pi t/P_i + \phi_i\right)$.

The unknown parameters are:
\begin{eqnarray}
\left\{N, \boldsymbol{\alpha}, \{\boldsymbol{\psi}\}_{i=1}^N, m_0,
\sigma_{\rm extra}, \nu\right\}
\end{eqnarray}
where $N$ is the number of planets,
$\boldsymbol{\alpha} = \{\mu_P, \sigma_P, \mu_A\}$ are hyperparameters
hyperparameters used to define the prior for the properties of the planets,
and $\psi_i = \{P_i, A_i, e_i, \phi_i, \omega_i\}$
are the properties of planet $i$.
The parameter $m_0$ describes a DC offset in the data, and
$\sigma_{\rm extra}$ and $\nu$ are parameters of the noise distribution
which are discussed further below. Note that our parameter $\omega_i$ is
standard, however $\phi_i$ is non-standard because we assert that $\phi_i=0$
always implies the signal is at its maximum at $t=0$. Our parameter space is
equivalent to the standard one, we are just using a different coordinate system.

A standard assumption for the probability distribution of the data given the
parameters (known as the sampling distribution, which becomes the likelihood
function when the dataset is known) is a normal distribution with
standard deviation $\sigma_i$ known from the error bars in the data set.
However, it is usually recommended to put in ``safety features'', in case the
data set contains any discrepant measurements, or in case the error bars in the
data set are underestimated. To achieve this, we used a student-$t$ distribution
instead of a normal distribution, with
scale parameter $\sqrt{\sigma_i^2 + \sigma_{\rm extra}^2}$ and shape parameter $\nu$. The
parameter $\sigma_{\rm extra}$ is an ``extra noise'' parameter that effectively increases the
size of the error bars, and the shape parameter $\nu$ allows for heavier tails
than a normal distribution. If $\nu$ is large, the student-$t$ distribution
is approximately a normal distribution, and if $\nu$ is small the noise
distribution has much heavier tails. For instance, when $\nu=1$ the student-$t$
distribution becomes a Cauchy distribution.

All of the model assumptions are specified in detail in
Table~\ref{tab:priors}. We assigned hierarchical priors to some of the planet's
parameters (i.e. the prior for the planets' parameters is defined conditional
on some hyperparameters). This allows the
model to capture the idea that knowing the values of some planet's parameters
provides some information about the parameters of another planet. Not using
hierarchical priors usually implies a strong prior commitment to the hypothesis
that the properties of the planets are spread out across the whole domain of
possible values, which is not necessarily the case.

Most of our priors were chosen to represent vague
prior knowledge, rather than the judgement of an informed expert on extrasolar
planets. Uniform distributions were used for parameters such as phases, where
time-translation symmetry seems plausible. For some parameters we assigned the
distribution in terms of the log of the parameter, rather than the parameter
itself, when the parameter is positive and uncertain by orders of magnitude.
Truncated Cauchy distributions were used when there is a
preferred value, but since these have very heavy tails, the assumption is
quite fail safe relative to other possible ``informative'' assignments such as normal distributions.
For example, the prior for $\mu_P$, the typical orbital period,
is centered around 1 year but could be as low as $\approx e^{-21}$ years or as high
as $\approx e^{21}$ years, a very generous range. A uniform distribution for
$\log(\mu_P)$ would have been more conventional, whereas the Cauchy distribution
expresses a slight preference for $\mu_P$ being of order one year.

An apparently strange choice is the conditional prior for the (logarithms of)
the orbital
periods, which is a biexponential distribution given a location parameter
$\mu_P$ and a scale parameter $w_P$ which determines the width of the
distribution\footnote{A biexponential distribution with location parameter
$\mu$ and scale parameter $w$ has probability density function
$p(x|\mu, w) = \frac{1}{2w} \exp\left(-\frac{|x - \mu|}{w}\right)$.}.
Rather than assigning independent priors to the log periods, the hierarchical
model allows for the periods to ``cluster around'' a typical period $\mu_P$ if
there is evidence for this. On the other hand, independent priors for the
periods would imply a strong prior commitment to the hypothesis that the
periods are spread out across the whole prior volume (equivalent to assuming
a fixed large value for $w_P$).
A more conventional choice for the conditional prior given $\mu_P$ and $w_P$
would have been a normal distribution. However, the \citet{rjobject} software
needs to know the corresponding cumulative distribution and its inverse, which
are not available in closed form for the normal distribution.
Our prior for $w_P$, which controls the diversity of the log-periods, was
uniform between 0.1 and 3, since it is unlikely that many planets have
extremely similar or extremely different (over several orders of magnitude)
orbital periods.

For the velocity semi-amplitudes $\{A_i\}$, we chose an exponential distribution
given the hyperparameter $\mu_A$ which sets the mean of the exponential distribution.
Our prior for $\mu_A$ spans many orders of
magnitude but expresses a slight preference for $\mu_A$ being of order unity,
using a Cauchy distribution. The prior for the semi-amplitudes
will influence how many of these low amplitude planets will be inferred: if we
believe there are many, and the data are uninformative about low amplitude
planets, then the posterior distribution for $N$ will also indicate that there
may be many low amplitude planets. However, their other properties, such as
their orbital periods, will not be well determined. The Beta prior for
eccentricity was suggested by Gregory (priv. comm) and is an approximation
to the inferred frequency distribution of eccentricities in the population
\citep{kipping}.

\begin{table*}
\begin{tabular}{|l|l|l|}
\hline
Quantity	&	Meaning		& Prior\\
\hline
{\bf Hyperparameters}	&	\\
$N$		& Number of planets	& Uniform$(\{0, 1, ..., N_{\rm max}\})$\\
$\mu_P$		&	Median orbital period (years)	& $\log(\mu_P) \sim$ Cauchy$({\rm location}=5.9, {\rm scale}=1)T(-15.3, 27.1)$\\
$w_P$		&	Diversity of orbital periods & $w_P \sim$ Uniform$({\rm min}=0.1, {\rm max}=3)$\\
$\mu_A$		&	Mean amplitude (metres per second)	& $\log(\mu_A) \sim$ Cauchy$({\rm location}=0, {\rm scale}=1)T(-21.2, 21.2)$\\
\hline
{\bf Planet Parameters}\\
$P_i$		&	Orbital period	&	$\log(P_i) \sim $ Biexponential$({\rm location}=\log(\mu_P), {\rm scale}=w_P$)\\
$A_i$		&	Semi-amplitude of signal	& Exponential$(\textnormal{mean}=\mu_A)$\\
$\phi_i$	&	Phase of signal	&	Uniform($\textnormal{min}=0, \textnormal{max}=2\pi$)\\
$e_i$		&	Eccentricity of orbit	&	Beta$(\alpha=1, \beta=3.1)T(0, 0.8)$\\
$\omega_i$	&	Viewing angle	&	Uniform$({\rm min}=0, {\rm max}=2\pi)$\\
\hline
{\bf Other}\\
$m_0$		&	Constant DC offset	&	Uniform$({\rm min}=m_{0, \rm min}, {\rm max}=m_{0, \rm max})$\\
$\sigma_{\rm extra}$	& ``Extra noise'' parameter	& $\log(\sigma) \sim$ Cauchy$({\rm location}=0, {\rm scale}=1)T(-21.2, 21.2)$\\
$\nu$		& Shape parameter for $t$-distribution for noise & $\log(\nu) \sim$ Uniform$({\rm min}=\log(0.01), {\rm max}=\log(1000))$\\
\hline
{\bf Data}\\
$Y_i$		& Radial velocity measurements	&
		Student-$t\left({\rm location}=m(t_i), {\rm scale}=\sqrt{\sigma_i^2 + \sigma_{\rm extra}^2}, {\rm shape}=\nu\right)$
\end{tabular}
\caption{All of the prior distributions in our Bayesian model.
The priors for the planet parameters are defined conditional on the values
of the hyperparameters. Uniform priors were used for parameters like phases.
For parameters where a rough initial guess is possible, heavy-tailed Cauchy
distributions were used so this information could be taken into account
in a non-dogmatic way. Time units are in days and amplitude units are
in metres per second. The maximum number of planets, $N_{\rm max}$, was set
to 10. The prior limits for $m_0$ were set to the minimum and maximum value
in the dataset, which is not strictly a valid strategy. Some of the distributions
were truncated for numerical reasons, the truncated intervals are specified with the
$T(a,b)$ notation.\label{tab:priors}}
\end{table*}

\section{Orbit Look-Up Table}\label{sec:orbits}
The expected (noise-free) signal due to an exoplanet is periodic, but
non-sinusoidal when the orbit is not perfectly circular. The expected
shape $m(t)$ of the variations is needed in order to evaluate the likelihood
function for any proposed setting of the parameters.
To save time, we pre-computed the properties of orbits as a function of
eccentricity. We also made the standard assumption that the planets do not
interact, so the expected signal due to several planets is the sum of the
contributions of each planet.

Consider a test particle moving in the $x$-$y$ plane under the influence of a
point mass at the origin. The motion of the test particle represents the
reflex motion of the host star orbiting around the center of mass of the
system. The equations of motion for the particle are:
\begin{eqnarray}
\frac{d^2x}{dt^2} &=& -\frac{x}{r^3} \\
\frac{d^2y}{dt^2} &=& -\frac{y}{r^3} \\
\end{eqnarray}
where $r = \sqrt{x^2 + y^2}$. The solutions to this system of equations are elliptical orbits
with the focus at the origin. We set the initial position to $(1, 0)$, and
the initial velocity to $(0, v)$ where $v \in [0.4, 1]$.
If $v=1$, the orbit is circular and as $v$ decreases the orbit becomes more
elliptical. For trial values of $v$ ranging from 0.4 to 1 in steps of 0.005,
we calculated the orbit, and saved the
velocities $\frac{dx}{dt}$ and $\frac{dy}{dt}$ as a function of time to disk.
These saved orbits were used as a lookup table for constructing the expected
signal $y(t)$ due to a single planet.
Because of the initial conditions, the simulated orbits were all horizontally
aligned. If the observer is located on the $x$-axis a large distance
from the origin, they will measure $m(t) = \dot{x}(t)$. However, if the
observer is located at an angle $\omega$ with respect to the $x$-axis, then
the radial velocity measured will instead be
$m(t) = \cos(\omega)\dot{x}(t) + \sin(\omega)\dot{y}(t)$.
Since the our orientation with respect to the orbits is unknown, each planet
requires a ``viewing angle'' parameter $\omega$ also known as the longitude
of the line of sight \citep{ohta}.
The eccentricity of the orbit, in terms of $v$, is $e = 1 - v^2$.
By precomputing a set of orbits before running the MCMC, we are able to
do $\sim$ 15,000 likelihood evaluations per second per CPU core.

\section{Demonstration on Simulated Data}\label{sec:fake_data}
\begin{figure}
\includegraphics[scale=0.4]{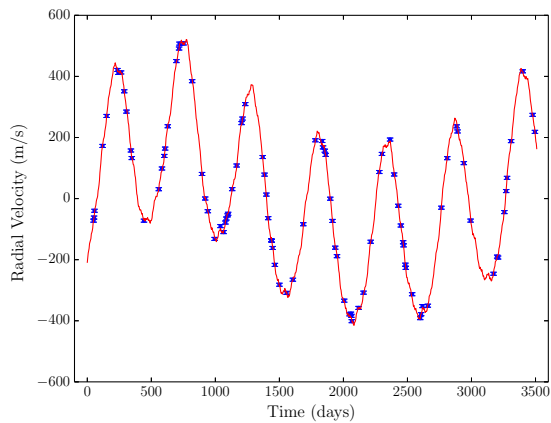}
\caption{A simulated dataset that ``resembles'' the $\nu$ Oph dataset, which
we used to test our methodology. The dominant signal is from two large
planets with periods periods of 530 and 3120 days and semi-amplitudes of
291 m s$^{-1}$ and 181 m s$^{-1}$ respectively. There are also five
much smaller planets which contribute small additional effects to the data.
\label{fig:fake_data}}
\end{figure}

To test our proposed methodology, we generated a simulated dataset for a
system with $N=7$ planets. The dataset was ``inspired by'' the $\nu$ Oph
dataset (Section~\ref{sec:nu_oph}), and contains two large signals with
periods of 530 and 3120 days, whose semi-amplitudes are
291 m s$^{-1}$ and 181 m s$^{-1}$ respectively. The other five planets have
much lower semi-amplitudes, ranging from 4-30 ms$^{-1}$. The standard deviation
for the noise in the data was 5 m s$^{-1}$, so some of these low-amplitude
signals should be detectable. The simulated data is shown in
Figure~\ref{fig:fake_data}, along with the true radial velocity curve $m(t)$
that was used to generate the data.

\begin{figure}
\includegraphics[scale=0.4]{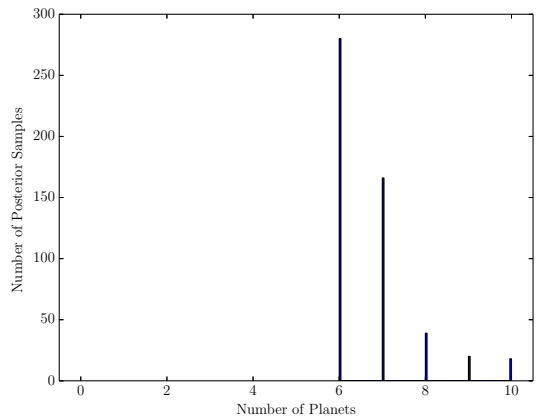}
\caption{The posterior distribution for the number of planets $N$ given the
simulated dataset. The true number of planets was 7.
\label{fig:fake_data_N}}
\end{figure}

We ran our algorithm on the simulated dataset to obtain samples from the
posterior distribution. We obtained 520 posterior samples.
The posterior distribution for $N$, the number of
planets, is shown in Figure~\ref{fig:fake_data_N}. The true number of planets,
7, is not the most probable value, but it does have substantial probability.
The posterior distribution suggests that $N$ could be anywhere from 6 to 10.

\begin{figure}
\includegraphics[scale=0.4]{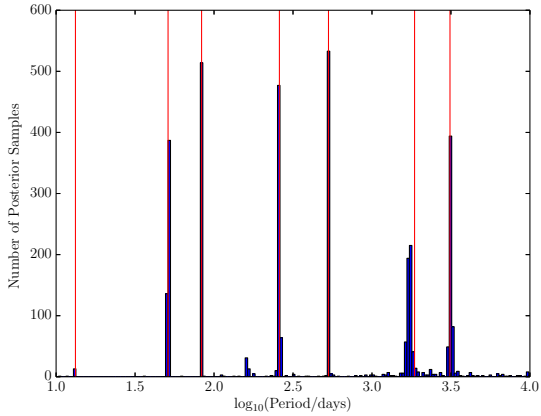}
\caption{The posterior distribution for the periods given the simulated
dataset. The solid lines are the true periods, six of which were detected
with high posterior probability.\label{fig:fake_data_periods}}
\end{figure}

The posterior distribution for the periods $\{P_i\}_{i=1}^N$ is shown in
Figure~\ref{fig:fake_data_periods}. Because of the label switching degeneracy,
the posterior distribution for each period is identical, so we pooled the
samples for all periods. Defining the log-periods by
$S_i =  \log_{10}\left[P_i/(\textnormal{1 day})\right]$,
Figure~\ref{fig:fake_data_periods} is a Monte Carlo representation
of the mixture distribution
\begin{eqnarray}
f(S) &=& \sum_{N=0}^{10} p(N | D)\sum_{i=1}^N p(S_i | N, D).
\end{eqnarray}
If a certain period is accurately measured (i.e. it appears in close to 100\% of
the posterior samples and the distribution for its period is very narrow)
then it will appear in Figure 3 with a height of $\sim 520$. If the uncertainty
in the period is larger than the histogram bin width then the peak will be
spread over several bins.

The posterior distribution for the periods, shown in
Figure~\ref{fig:fake_data_periods}, shows that six of the true periods were
recovered, with probability close to 1. One period (with $\log_{10}P_i \approx 1.15$)
which was actually present was not ``detected'' because it had a very small
amplitude. We note that the posterior probability near this period should not be
precisely zero. There is also some
evidence for periods which did not actually exist, however the posterior
probabilities for these peaks are not close to 1.

\begin{figure}
\includegraphics[scale=0.4]{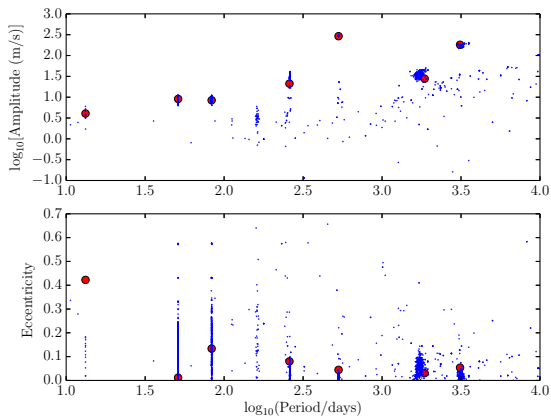}
\caption{The joint posterior distribution for the periods and the amplitudes
(top panel) and the periods and eccentricities (bottom panel) from the
simulated dataset. The true values are plotted as circles.
\label{fig:fake_data_posterior}}
\end{figure}

The joint posterior distribution for the periods and the amplitudes of the
signals is shown in Figure~\ref{fig:fake_data_posterior} along with the
eccentricities. As with Figure~\ref{fig:fake_data_periods}, the samples for
all planets were combined. The true values are also plotted as circles. Clearly, the
reason the period of $\log_{10}(P_i) \approx 1.15$ was not ``detected'' was that
it had a very low amplitude of approximately 4 m s$^{-1}$ which is below the
noise level.

\section{Application to $\nu$ Oph}\label{sec:nu_oph}
\begin{figure}
\includegraphics[scale=0.45]{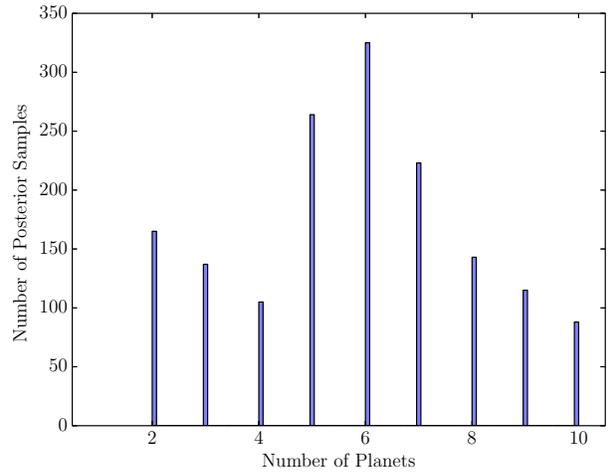}
\caption{The posterior distribution for the number of planets $N$ orbiting
$\nu$ Oph. The posterior probability of $N > 2$ is about 88\%, however the
prior probability of $N > 2$ was already high due to the uniform prior for $N$.
\label{fig:nu_oph_N}}
\end{figure}

The $\nu$ Oph system is generally accepted to have two confirmed planets
\citep[e.g.][]{2011AIPC.1331..102Q, 2012PASJ...64..135S, fengji}, with periods
of $530.3$ and $3190$ days. To test our approach we applied it to the RV
data from \citet{2012PASJ...64..135S}.
The posterior distribution for $N$, the number of planets, is shown in
Figure~\ref{fig:nu_oph_N}, showing that $N$ could be anywhere from 2 to 10,
and the posterior probability that $N \geq 3$ is about 88\%. Of course, these
extra possible signals are not necessarily a planet but a feature in the data
which is better explained by a periodic signal than by noise (and may have been
explained by correlated noise, had we included it).

\begin{figure}
\includegraphics[scale=0.45]{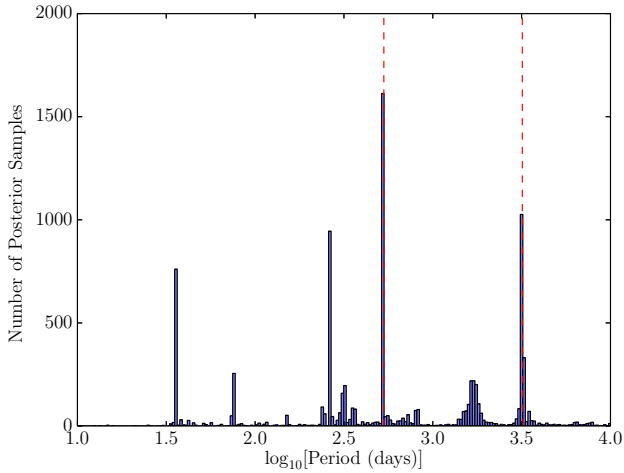}
\caption{The posterior distribution for the orbital periods in the $\nu$ Oph
system. The two dashed vertical lines are the commonly accepted periods
of $530.3$ and $3190$ days. The next most prominent peaks with well determined
periods are at log periods of around 1.6, 2.4, and 3.3, corresponding to
periods of 36.11 $\pm$ 0.034 days,
75.58 $\pm$ 0.80 days, and 1709 $\pm$ 183 days.\label{fig:nu_oph_periods}}
\end{figure}

\begin{figure}
\includegraphics[scale=0.45]{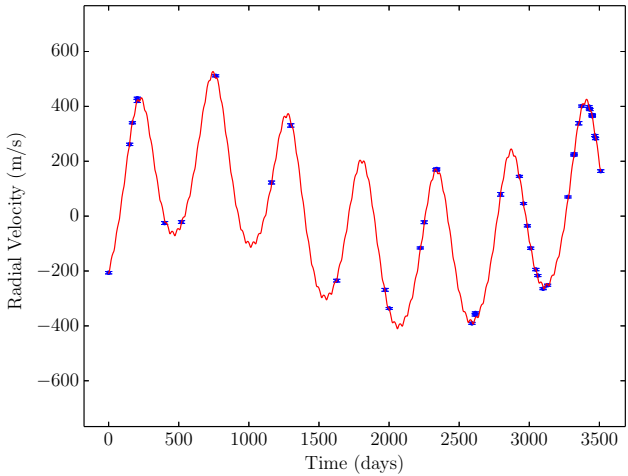}
\caption{The $\nu$ Oph radial velocity data and an example model fit which
includes a third period. The amplitude of this additional signal is low but
is about twice the reported errorbars on the measurements.\label{fig:nuoph}}
\end{figure}

The posterior distribution for the logarithms of the periods is shown in
Figure~\ref{fig:nu_oph_periods}. As in Section~\ref{sec:fake_data},
the posterior samples for all periods were combined to make this figure, which
shows several prominent peaks. 
The two peaks with vertical dashed lines are the commonly accepted periods of
530 and 3190 days, and the other prominent peaks (i.e. signals which have a
moderate probability of existence) have periods of 36.11 $\pm$ 0.034 days,
75.58 $\pm$ 0.80 days, and 1709 $\pm$ 183 days. As with any MCMC output,
if we are interested in the probability of any proposition $Q$ (for example,
``$Q \equiv$ a planet exists with period between 35 and 37 days''), we can
calculate the proportion of the posterior samples for which $Q$ is true, which
(if we have a lot of samples) is a Monte Carlo estimate of the posterior
probability of $Q$. For $\nu$ Oph, we calculated the probability that
at least one of these ``extra'' signals (beyond the two commonly accepted ones)
exists, as 85\%.
Given that they exist, their amplitudes are low, around 5-40 metres per second,
which we note is above the noise level. An example model fit to the data is
shown in Figure~\ref{fig:nuoph}.

\begin{figure}
\includegraphics[scale=0.45]{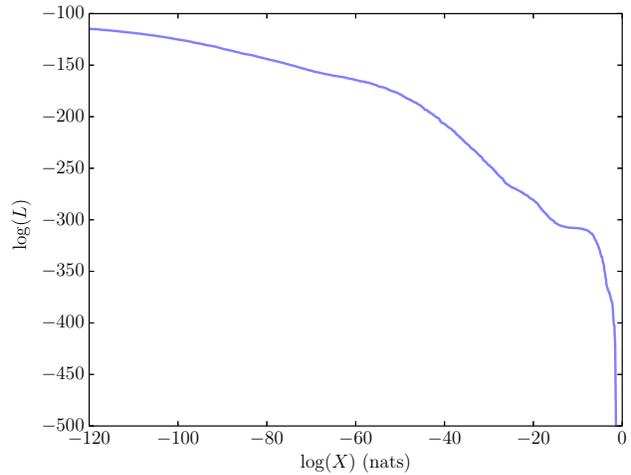}
\caption{Log likelihood vs. enclosed prior mass (a standard output of Nested
Sampling algorithms) for the $\nu$ Oph analysis.
There are several phase transitions (concave-up regions) present. The small one at
$\log(X) \approx -70$ separates models which contain additional signals from
models that do not. Without using Nested Sampling it would be more difficult
to mix between these two situations and calculate the posterior probability
for the existence of the additional signals.
\label{fig:logl0}}
\end{figure}

Figure~\ref{fig:logl0} shows the relationship between the likelihood $L$
and the enclosed prior mass $X$ for the $\nu$ Oph analysis. These plots are
a standard output of Nested Sampling analyses, and provide
insights into the structure of the problem. Concave-up regions of this curve
indicate phase transitions which can cause severe problems for annealing-based
methods, and sometimes even for sampling the posterior distribution. In this
analysis, the models with a third signal exist to the left of the phase
transition at $\log(X) \approx -70$, and models without the signal exist to the
right of the phase transition. Mixing between these two phases is crucial for
accurately computing the posterior probability that extra signals exist.

The marginal likelihood for our model was
$\log(\mathcal{Z}) \approx -220.5$. Nested Sampling also allows for calculation
of the ``information'', or Kullback-Leibler divergence (a quantity in information
theory) from the prior to the posterior, which quantifies how much we learned about the parameters:
\begin{eqnarray}
\mathcal{H} &=& \int p(\theta | D) \log\left[\frac{p(\theta | D)}{p(\theta)}\right] \, d\theta
\end{eqnarray}
An intuitive interpretation of this quantity is the number of times the
prior distribution had to be compressed by a factor of $e$ (if the logarithm
in the formula is a natural logarithm) to get to the posterior distribution.
For the $\nu$ Oph data the information was $\mathcal{H} \approx 76.6$ nats
(natural units) or 111 bits, so the posterior occupies about $e^{-76.6}$ times
the prior volume.

\section{Application to Gliese 581}
The red dwarf star Gliese 581 is thought to host several planets. Exactly
how many is a matter of considerable debate. According to
\citet{2014Sci...345..440R}, there are two planets
(b and c, with periods 5.36 and 12.91 days respectively)
whose existence is generally
accepted, two more
(d and e, with periods 66 and 3.15 days respectively)
whose existence was mostly accepted, and another two
(f and g, with periods 433 and 36.5 days respectively)
whose existence was generally doubted. However, \citet{2014Sci...345..440R}
found that planets d and g do not exist but are signals due to stellar activity.
While our model cannot account for stellar variability and contribute to that
particular discussion, it is a challenging and interesting dataset from an
inference point of view.
To run our code on the combined dataset from the HARPS and HIRES spectrographs,
we extended the model to include separate DC offsets for each instrument, as
well as separate ``extra noise'' parameters $s_0$ and $\nu$. We also increased
$N_{\rm max}$ from 10 to 15 for this system.

\begin{figure}
\includegraphics[scale=0.45]{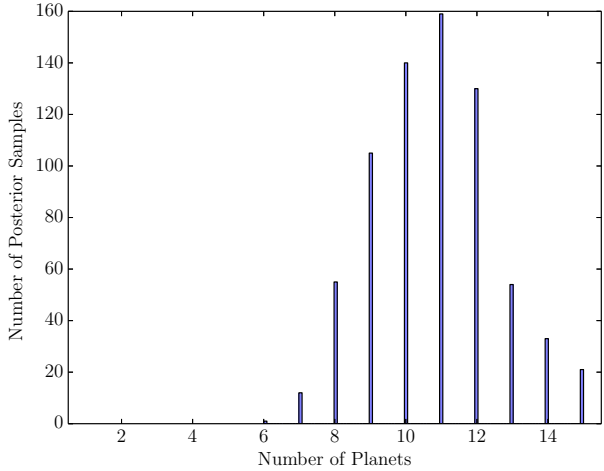}
\caption{The posterior distribution for the number of planets $N$ orbiting
Gliese 581.\label{fig:gliese581_N}}
\end{figure}

The posterior distribution for $N$ is shown in Figure~\ref{fig:gliese581_N},
and shows strong evidence for at least eight periods
($P(N \leq 7 | D) = 0.012$). Some authors \citep[e.g.][]{2011A&A...528L...5T}
recommend that the probability of $N$ planets should be $\geq$ 150 times
greater than the probability of $N-1$ planets existing before making a
claim that $N$ planets have been definitively detected. Such a decision rule
is presumably equivalent to a utility function where false positives are much
worse than false negatives. We note that applying this rule to our results, we
would assert $N=6$, even though this has a very small posterior probability.

\begin{figure}
\includegraphics[scale=0.45]{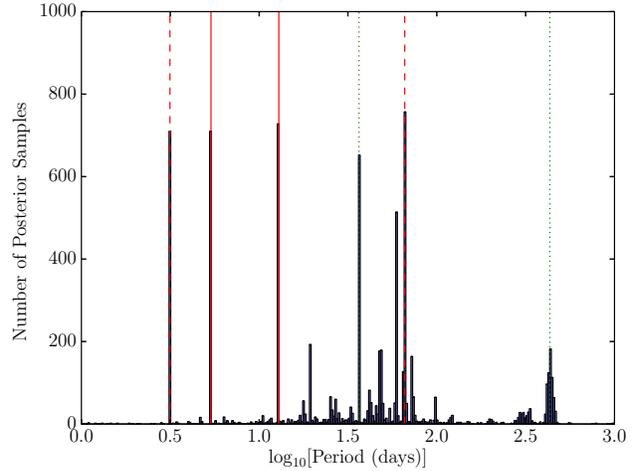}
\caption{The posterior distribution for the orbital periods in the Gliese 581
system. The solid lines are Gliese 581 b and c, the dashed lines are
``planets'' d and e, and the dotted lines are f and g.
\label{fig:gliese581_periods}}
\end{figure}

It is now recognised that
there are many possible sources for oscillations in a data set and not all
such oscillations should be claimed as planets. 
Our model cannot distinguish between oscillations due to planets and
oscillations due to stellar activity: any oscillations found in the dataset
will be described as ``planets'' by the model. Another important consideration
is the physical stability of the orbital system which is ignored by this type
of analysis \citep{orbital}. However it is interesting that
we find many more signals in the data than previous authors.
By inspecting the posterior distribution for the periods
(Figure~\ref{fig:gliese581_periods}), we see that only 4 of the periods
are well determined and have a posterior probability close to 1 (i.e. they
are present in all samples), corresponding to the known periods of Gliese 581
b, c, d, and e. The other ``periods'' are more uncertain.
As with $\nu$ Oph, we can
calculate the posterior probability of any hypothesis
about Gliese 581 by computing the fraction of the posterior samples that have
that property. The posterior probabilities for planets b, c, d, and e, are
close to 1. The posterior probability a signal exists with $\log_{10}(P)
\in [1.55, 1.57]$ is 88\%, and the probability for a signal with
$\log_{10}(P) \in [2.6, 2.8]$ is 85\%.

\begin{figure}
\includegraphics[scale=0.45]{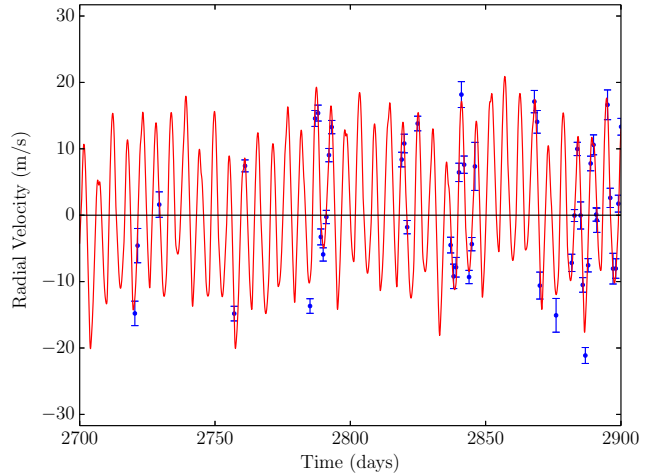}
\caption{A portion of the Gliese 581 data and an example model fit.
\label{fig:gliese581}}
\end{figure}

One possible explanation for the large number of inferred signals that a
non-sinusoidal signal due to stellar activity is being modelled as several
periods \citep[e.g. as happens in asteroseismology when sinusoidal models are
used;][]{astero}, and if the model were extended to include a ``stochastic''
oscillation \citep[e.g.][]{gaussproc}, the number of periods detected may be reduced
substantially. Another contributing factor is the prior for the amplitudes.
With these kinds of models, the posterior distribution for $N$
can be influenced by the prior for the amplitudes $A_i$. Many authors
assign independent broad priors to the amplitudes, and this causes the
``Occam's razor'' penalty for adding extra signals to be quite strong. Since we
use a hierarchical prior for the amplitudes, if some amplitudes are found to
be low, $\mu_A$ will become small. When $\mu_A$ is small, it is likely that any
extra signals will have small amplitudes, so the ``Occam's razor'' effect is
weaker. An example model fit for Gliese 581 is shown in
Figure~\ref{fig:gliese581}.

\begin{figure}
\includegraphics[scale=0.45]{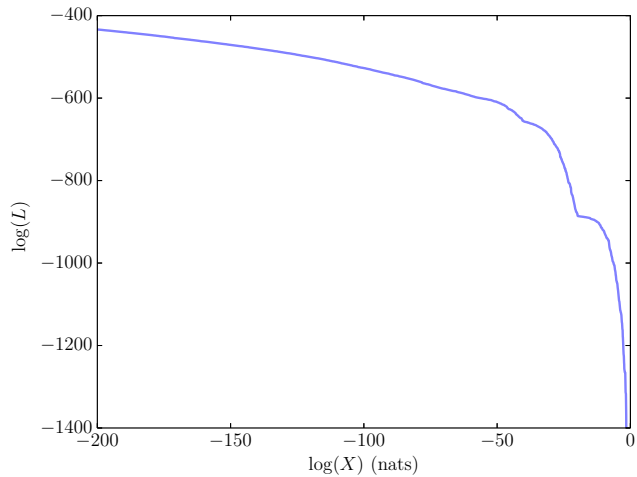}
\caption{Log likelihood vs. enclosed prior mass for the Gliese 581 analysis.
The two concave-up regions (at $\log(X) \approx -25$ and -45) correspond to
phase transitions. Thermal approaches to this problem would produce misleading
estimates of the marginal likelihood because they would mix poorly at temperatures
around 11 and 4.
\label{fig:logl}}
\end{figure}

The marginal likelihood was $\log(\mathcal{Z}) \approx -616.3$ and the
information was $\mathcal{H} \approx 130.4$ nats. This compares favourably
to the marginal likelihood of $-640.1$ (for a 6-planet model)
found by \citet{fengji}, although it is unclear whether we used exactly the
same dataset.
Interestingly, the log-likelihood curve (Figure~\ref{fig:logl})
shows this problem has two phase transitions. While these do not affect the
posterior distribution (as they did for $\nu$ Oph), they would cause difficulties
if we tried to calculte the marginal likelihood using annealing.

\section{Conclusions}
In this paper we introduced a trans-dimensional MCMC approach to inferring
the number of planets $N$ in an exoplanetary system from radial velocity data.
The MCMC was implemented using the framework of \citet{rjobject} which
defines trans-dimensional birth and death moves, and does the sampling
with respect to a Nested Sampling target distribution, rather than directly
sampling the posterior. This approach allows us to compute the results in a
single run, which provides posterior samples and an estimate of the marginal
likelihood. By using Diffusive Nested Sampling, instead of
directly trying to sample the posterior distribution, we can overcome difficult
features in the problem, such as phase transitions and (to some extent)
multiple modes.

We applied the code to two well-studied RV datasets, $\nu$ Oph and Gliese 581.
In $\nu$ Oph, we found some evidence for additional signals with low amplitude,
but with several possible solutions for their periods. Given our modelling
assumptions, the posterior probability at least one of these additional signals
is real is 85\%. The posterior distribution contains models both with and without
these additional signals, however, these are separated by a phase transition.
Therefore mixing between the two situations would be infrequent if we simply
tried to sample the posterior distribution.

With the combined HIRES+HARPS dataset from Gliese 581, we found
evidence for a large number of ``planets'', although only four have well
determined periods, corresponding to the Gliese 581 b, c, d,
and e. Since our model does not include any possibility of stellar variability,
any such periodic signals will be attributed to ``planets''.
Including non-planetary stellar variability is a crucial next step.

\vspace{-0.5cm}
\section*{Acknowledgements}
It is a pleasure to thank Fengji Hou and David Hogg (NYU) for inspiring me (BJB) to
finally work on this problem, and Phil Gregory (UBC) for writing so many
interesting papers on it. We also thank Tom Loredo (Cornell) and Dan
Foreman-Mackey (NYU) for interesting conversations and feedback, and Ben
Montet (Caltech) and Geraint Lewis (Sydney) for helpful discussions.
The referee also provided excellent suggestions for improving the manuscript.

This work was supported by a Marsden Fast Start
grant from the Royal Society of
New Zealand.

\end{document}